\documentclass[aps,pre,groupedaddress,notitlepage]{revtex4-1}
\usepackage{amsmath,amssymb} 
\usepackage[utf8]{inputenc} 
\usepackage[english]{babel} 
\usepackage{graphicx} 
\usepackage{listings} 
\usepackage{soul}
\usepackage{color}
\usepackage{subcaption} 
\usepackage{caption}
\usepackage{hyperref}
\hypersetup{
    colorlinks=true, 
    linktoc=all,     
    linkcolor=blue,  
}
\usepackage{xr}
\begin{document}

\title{The microscopic role of deformation in the dynamics of soft colloids }
\author{Nicoletta Gnan}\email{nicoletta.gnan@roma.infn.it}
\author{Emanuela Zaccarelli}\email{emanuela.zaccarelli@cnr.it}
\affiliation{CNR-ISC (National Research Council - Institute for Complex Systems) and Department of Physics, Sapienza University of Rome, Piazzale A. Moro $2$, $00185$ Roma, Italy}
\date{\today}
\maketitle

{\bf Soft colloids allow to explore high density states well beyond random close packing. An important open question is whether  softness controls the dynamics under these dense conditions. While experimental works reported conflicting results, numerical studies so far mostly focused on simple models that allow particles to overlap, but neglect particle deformations, thus making the concept of softness in simulations and in experiments very different. To fill this gap here we propose a new model system consisting of polymer rings with internal elasticity. At high packing fractions the system displays a compressed exponential decay of the intermediate scattering functions and a super-diffusive behavior of the mean-squared displacement. These intriguing features are explained in terms of the complex interplay between particle deformations and dynamic heterogeneities, which gives rise to persistent motion of ballistic particles. We also observe a striking variation of the relaxation times with increasing particle softness clearly demonstrating the crucial role of deformation in the dynamics of realistic soft colloids.}

In recent years, colloidal particles have emerged as useful models that give access to phases and states with no counterpart in atomic and molecular systems~\cite{ ruzicka2011observation,dotera2014mosaic,chen2011directed}. In addition they have allowed to establish new mechanisms to control phase behaviour~\cite{hertlein2008direct, sacanna2011lock, van2013entropically} and to deepen our understanding of the glass and jamming transition~\cite{pham2002multiple, royall2008direct, mattsson2009soft}. 
A crucial parameter controlling colloidal behaviour is particle softness, which can be quantified by the ratio between  elastic  and thermal  energy~\cite{vlassopoulos2014tunable}. Hence, particle internal elasticity is the key ingredient to distinguish hard particles like sterically stabilized polymethylmethacrylate (PMMA) colloids from soft and ultrasoft ones such as microgels, emulsions or star polymers to name a few.  Several experimental works~\cite{mattsson2009soft,seekell2015relationship,nigro2017dynamical,nigro2018structural} reported that softness controls the dependence of the structural relaxation time $\tau_{\alpha}$ on temperature  $T$ or on packing fraction $\zeta$ -- the so called fragility\cite{angell1995formation}.  A system is called fragile when the $\tau_{\alpha}$ dependence is described by a Vogel-Fulcher-Tamman law~\cite{debenedetti2001supercooled}, meaning that its variation is large over small changes of $T$ or $\zeta$; contrarily, strong systems are characterized by an Arrhenius behaviour, implying a mild variation of $\tau_{\alpha}$ upon varying the control parameter.
While the pioneering study of Mattsson and coworkers~\cite{mattsson2009soft} proposed a link between elasticity and fragility, there is still no consensus on this issue. A recent work based on a simple theoretical model has confirmed that such a link exists~\cite{van2017fragility}, but this picture has been later challenged by experiments on colloids of different softness~\cite{philippe2018glass}. To gain microscopic knowledge on this matter, we usually resort on simulations of simple repulsive models, as for example systems interacting with the Hertzian potential~\cite{landau1986theory}, which is found to describe microgel particles behavior at moderate packing fractions~\cite{mohanty2014effective,articoloMaxime}, but is expected to fail in denser conditions where soft colloids tend to shrink, deform or even interpenetrate~\cite{mohanty2017interpenetration}. 
Early works have indicated that, for such simple pair potentials, the change of dynamic properties with softness, such as the change of fragility, is modest~\cite{sengupta2011dependence} or absent ~\cite{de2004scaling}. 
In these approaches, softness is tuned by modifying a given parameter, e.g. the strength of the repulsion, allowing particles to overlap to a certain extent, but without taking into account their deformability as well as other important aspects in realistic soft particles, namely deswelling~\cite{urich2016swelling,de2017deswelling,weyer2018concentration,higler2018apparent}, interpenetration~\cite{mohanty2017interpenetration} and faceting~\cite{conley2017jamming}. Thus, there is a strong need to go one step forward in the modeling of soft colloids to tackle this problem and to provide a microscopic picture of these systems at high densities.

\begin{figure}[!htb]
\centering
\includegraphics[width=1.0\textwidth]{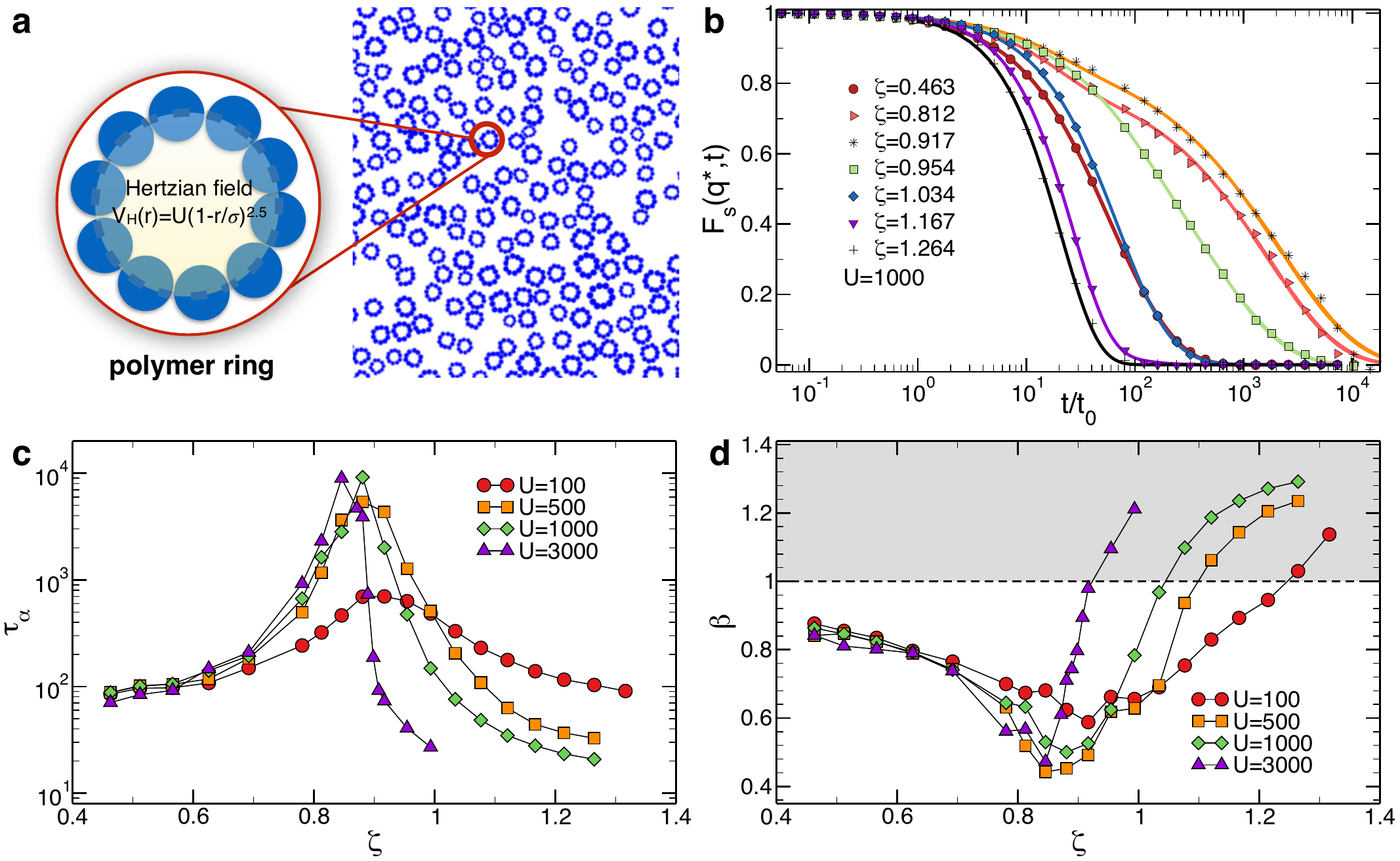}
\centering
\caption{{\bf Model system and dynamical properties as a function of packing fraction $\zeta$}: (a) snapshot of a portion of EPR with $U=1000$ and $\zeta=0.463$ and illustration of the model. Representative snapshots at different $\zeta$ values are reported in Fig.~S1;
(b) self-intermediate scattering function $F_{s}(q^{*},t)$ for EPR with $U=1000$. Points are simulation data and solid lines are fits using Eq.~\ref{eq:fitbeta};  (c, d) relaxation time $\tau_{\alpha}$ and shape parameter $\beta$ extracted from the fits to $F_{s}(q^{*},t)$ for rings with different $U$. In panel (d), the grey area highlights the compressed exponential region.}
\label{fig:Panel1}
\end{figure}

To try to reconcile experimental and numerical results, in this work we investigate a new model of elastic polymer rings (EPR) that explicitly shrink and deform. Inspired by recent experiments on ultrasoft microgels with tunable internal elasticity~\cite{Bachman_2015, Virtanen_2016}, we add a Hertzian potential of repulsive strength $U$ (Eq.~\ref{eq:hertzian}) in the centre of mass of polymer rings (see Fig.~\ref{fig:Panel1}(a)). This term allows the rings to retain a circular shape at low $\zeta$ but also provides an energetic cost associated to particle deformation. We perform 2D extensive numerical simulations of polydisperse rings upon varying $U$ for a wide range of $\zeta$ up to very dense states, where faceting effects become important (see Fig.~S1 and Supplementary Movie).  More details on the model are given in the Methods section.  

We report in Fig.~\ref{fig:Panel1}(b) the self-intermediate scattering functions $F_{s}(q^{*},t)$ at the wavenumber $q^{*}$, corresponding to the maximum of the static structure factor, for different packing fractions $\zeta$ and a fixed value of the amplitude of the Hertzian potential $U=1000$. The associated relaxation time $\tau_{\alpha}$, shown in Fig.~\ref{fig:Panel1}(c), at first increases, indicating a slowing down of the dynamics up to $\zeta=\zeta_R\approx 0.9$. Above this packing fraction, the system becomes faster, due to melting upon compression, for all studied $U$ values. Such a reentrant behavior was already observed for 3D Hertzian spheres~\cite{berthier2010increasing} and in simulations of single-chain nanoparticles~\cite{verso2016tunable}, and here confirmed for 2D Hertzian disks (HZD), as reported in the SI.
Although reentrant melting occurs both in EPR and in HZD, its mechanism is very different in the two cases: HZD do this by overlapping, while EPR through particle deformation which is accompanied by the accumulation of internal stresses. This difference is encoded in the shape of $F_{s}(q^{*},t)$, which is described by the shape parameter $\beta$ defined in Eq.~\ref{eq:fitbeta}: the decay of $F_{s}(q^{*},t)$ has always a stretched exponential form ($\beta <1$) for HZD as also shown in Fig.~S2, while for EPR  it becomes faster than exponential ($\beta >1$) for $\zeta\gtrsim \zeta_R$, as shown in Fig.\ref{fig:Panel1}(d). This signals the onset of a compressed exponential relaxation of the density auto-correlation functions that is found for all studied $U$ (see Fig.~S3) with $\beta$ values strongly dependent on softness and on $\zeta$.

\begin{figure}[h!]
\hspace{-0.04cm}
\centering
\includegraphics[width=1.0\textwidth]{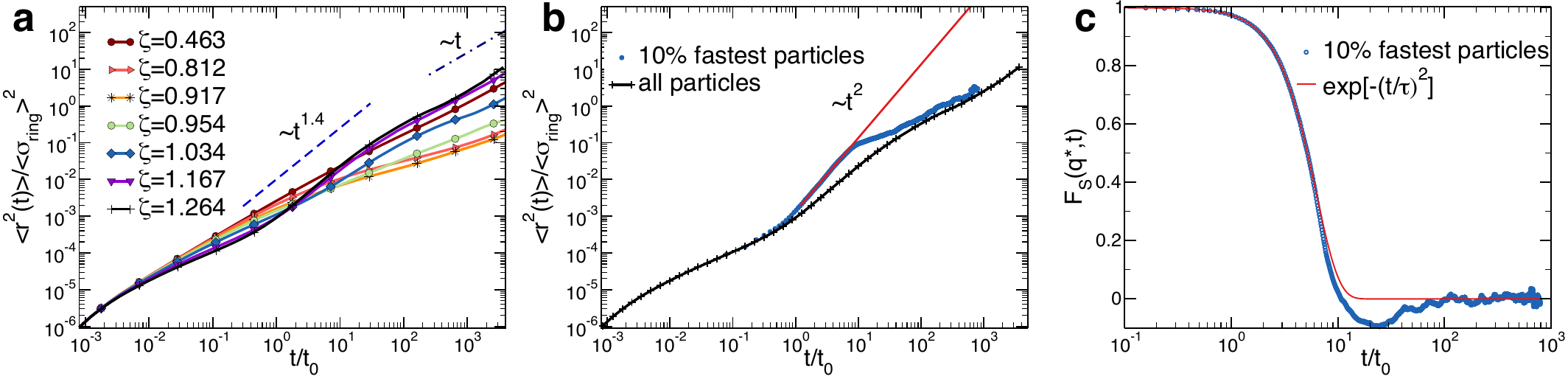}
\centering
\caption[c]{{\bf Mean-squared displacement and ballistic particles}: (a) MSD $\langle r^2(t)\rangle$ for EPR with $U=1000$ at different $\zeta$. Dashed lines are guides to the eye to show super-diffusive behavior $\langle r^2(t)\rangle \sim t^{\gamma}$ and normal diffusion at long times; (b) MSD of all particles (solid black line with symbols) and of ballistic fastest particles (blue symbols), detected in a time interval $\Delta t=7.88$ in reduced units,  for $U=1000$ and $\zeta=1.264$. The solid line is a power-law fit to the data yielding $\gamma=2.003\pm 0.004$; (c) $F_{s}(q^*,t)$ for ballistic fastest particles (symbols).  The solid line is a compressed exponential fit to the data yielding $\beta=2.019 \pm 0.012$.}
\label{fig:Panel2}
\end{figure}

The compressed exponential decay of $F_{s}(q^{*},t)$ for $\zeta \gtrsim \zeta_R$ is accompanied by a super-diffusive behavior of the mean-squared displacement (MSD), i.e. $\langle r^2 (t) \rangle \sim t^{\gamma}$ with $\gamma >1$, as shown in Fig.~\ref{fig:Panel2}(a). This holds in an intermediate time window of about two decades, while, at long times, the MSD always becomes diffusive again. Similar results are found for all studied values of $U$ with the exponent $\gamma$ strongly depending on $\zeta$ and $U$ (see Fig.~S4) in analogy with the shape parameter $\beta$.

To grasp the microscopic origin of the observed compressed/super-diffusive behavior and the associated exponents, we analyse the system  in terms of dynamic heterogeneities. To this aim we monitor the dynamics of fast particles at high $\zeta$ in a time window $\Delta t$ of the order of $\tau_{\alpha}$ (see Methods). We find that the MSD of fast particles, averaged over several time windows, 
displays a super-diffusive behavior such as the total one, albeit with a significantly higher exponent. Thus, we separately analyse different time windows and find that for a large number of them (see Methods), the MSD of the fastest particles obeys a purely ballistic dynamics, i.e. $\langle r^2(t)\rangle \sim t^2$  in the considered time interval, as shown in Fig.~\ref{fig:Panel2}(b). At long times, still they recover diffusive behavior.

Further insights can be gained by looking at  $F_{s}(q^*,t)$ for the same selected fastest particles, which is shown in Fig.~\ref{fig:Panel2}(c)): it displays a compressed exponential decay with exponent $\beta=2$ in the same time window. 
We also carried out a similar analysis for different values of $\zeta$ and $U$, finding that these features are preserved but the amount of fastest particles showing ballistic behavior varies, in particular it increases with $\zeta$ and $U$. This allows us to explain the observed anomalous dynamics in terms of a superposition of different particle populations, including groups of ballistic particles whose size 
depends on softness and packing fraction, reflecting the increase of the exponents $\beta$ and $\gamma$ with $U$ and $\zeta$. 

Interestingly, at long times, $F_{s}(q^*,t)$ becomes negative (see Fig.~\ref{fig:Panel2}(c)) before eventually decaying to zero. This intriguing behavior has also been observed in active particles~\cite{kurzthaler2016intermediate,schwarz2016escherichia} and is the signature of a persistent motion of particles in a preferential direction. Indeed, it can be shown~\cite{berne2000dynamic} for non-interacting ballistic particles moving with velocity $v$ in the same direction that $F_{s}(q,t)\sim J_{0}(qvt)$,  i.e. a Bessel function of zero-th order.

The compressed exponential relaxation is still an important open question in colloidal systems and glass-formers~\cite{cipelletti2000universal,angelini2013dichotomic,ruta2013compressed}. 
Previous works on colloidal gels have linked the presence of such feature to the accumulation of local stresses which are then released into the system, triggering the faster-than-exponential/diffusive dynamics~\cite{cipelletti2000universal, bouchaud2002anomalous, cipelletti2003universal,duri2006length}. Recent simulations investigated this process 
by artificially altering the network dynamics~\cite{bouzid2017elastically} to observe stress propagation into the system. However, no evidence has been provided so far of compressed exponential relaxation in a microscopic elastic model undergoing spontaneous relaxation.

\begin{figure}[h!]
\hspace{-0.04cm}
\centering
\includegraphics[width=1.0\textwidth]{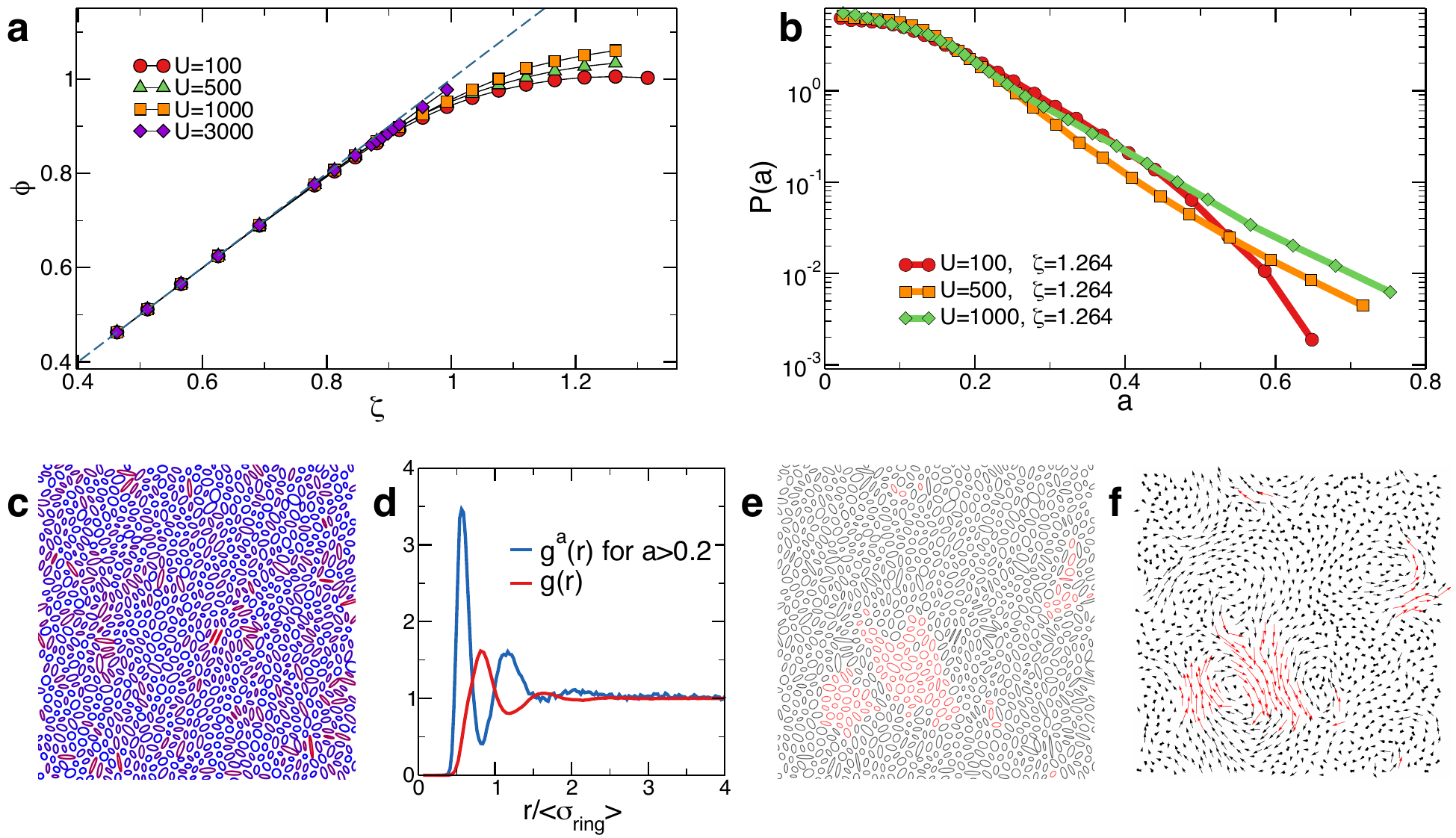}
\centering
 \caption{{\bf Analysis of rings deformation}: (a) Effective packing fraction $\phi$ vs nominal packing fraction $\zeta$ for all investigated $U$ values; (b) $P(a)$ for $\zeta=1.264$ and different $U$ values; (c) Snapshot of EPR represented as ellipses whose semi-axes correspond to the eigenvectors of the gyration tensor; (d) radial distribution functions $g(r)$ (for all rings) and $g^{a}(r)$ (for strongly deformed rings with $a>0.2$)  for $U=1000$ and $\zeta=1.264$;  (e) Same snapshot as in (d) highlighting in red the ballistic fastest particles in a time interval $\Delta t=7.88$ in reduced units and (f) the associated displacements in the same time interval. Clearly, fast ballistic particles show persistent, correlated motion within large groups. The displacements are magnified (by an arbitrary factor) in order to improve visualization.}
\label{fig:Panel3}
\end{figure}

By taking into account particle deformation in our model we are now able 
to quantify the local stress and to connect it to the onset of the compressed exponential behavior. To this aim we define the asphericity parameter $a$ (see Methods) that describes the deviation of the ring shape from a circular one: larger values of $a$ thus correspond to more deformed particles.  The distributions of particle asphericity $P(a)$ (see Fig.~S5) indicate that upon increasing $\zeta$ a larger and larger fraction of particles undergoes a strong deformation. A direct link exists between particle deformation and intra-ring stress, as discussed in the SI (see Fig.~S6). In order to quantify the effect of deformations at high $\zeta$, we calculate the effective packing fraction $\phi$ occupied by the rings (see Methods), which is shown in Fig.~\ref{fig:Panel3}(a).
We find that $\phi$ coincides with the nominal packing fraction $\zeta$ for $\zeta\lesssim \zeta_R$, while it becomes significantly smaller than $\zeta$  for denser states. Softer particles are found to deviate earlier from the linear $\phi-\zeta$ relation than stiffer particles, as observed in experiments for ionic microgels~\cite{pelaez2015impact}. Finally, for very large $\zeta$, a strong bending of $\phi$ is found, resulting even in a non-monotonic behavior for the softest rings. These findings clearly indicate that our simple model is able to capture another of the main ingredients of realistic soft particles, i.e. deswelling at high concentrations (due to shape deformation).

We now examine the variation of $P(a)$ with softness in Fig.~\ref{fig:Panel3}(b): while one may have expected to find stronger deformations in softer rings, this turns out not to be the case. Indeed, stiffer rings are found to display a longer tail for large $a$ values. This is counter-acted by the fact that, for a wide range of intermediate $a$ values, $P(a)$ is larger for softer rings (e.g. $U=100$). Thus, soft EPR prefer to undergo a large spread of moderate deformations avoiding too high $a$ values, while stiffer ones tend to accumulate strong deformations within a small fraction of rings (e.g. $U=500$). When $U$ grows even further, also intermediate deformations grow (e.g. $U=1000$).   
To understand the correlation between deformation and dynamics, we calculate the auto-correlation function of the asphericity, and thus indirectly of the stress, in Fig~S7, finding that its relaxation time $\tau_{asph}$ roughly coincides with that at which the super-diffusive behavior terminates. This result clearly connects the stress-releasing and stress-building mechanism with the occurrence of ballistic motion: within $\tau_{asph}$, deformed particles release stress (reducing their asphericity) and contribute, with all other particles, in triggering the motion of less stressed particles.  Since this is a mechanism involving the whole system, it is difficult to completely isolate each contribution, but our findings strongly suggest that stress-governing events entirely control the onset of the ballistic motion and the associated compressed exponential relaxation. Thus, in analogy with colloidal gels, the underlying physical mechanism of the compressed/super-diffusive dynamics is stress propagation, that is spontaneously obtained within our model through particle deformation.

To visualize how deformation and stress are distributed within the system, a snapshot of EPR with $U=1000$ at the highest investigated $\zeta$ is reported in Fig.~\ref{fig:Panel3}(c), where each ring is represented by an ellipse based on the eigenvectors of its gyration tensor. It is evident that deformed particles tend to stay close to each other, generating "strings" of elongated ellipses, which surround areas of less deformed/less stressed particles. This is quantified by the radial distribution function $g^{a}(r)$ of rings with large asphericity (i.e. $a>0.2$), which displays a higher peak located at smaller distances with respect to the average $g(r)$,  as shown in the inset of Fig. \ref{fig:Panel3}(d). 
The same snapshot is also shown in Fig.~\ref{fig:Panel3}(e) highlighting the fastest ballistic rings, which are remarkably found in very large clusters. Furthermore, not only their positions are correlated, but also their displacements, as shown in Fig.~\ref{fig:Panel3}(f): a high degree of alignment in the direction of motion is clearly present, in full agreement with the observation of a negative oscillation in $F_s(q^*,t)$ (Fig.~\ref{fig:Panel2}(c)). On the boundary of these clusters, slower rings which are either more deformed or are moving in a different direction physically stop the ballistic motion of the fastest particles which are then slowed in their motion until recovering normal diffusion (Fig.~\ref{fig:Panel2}(b)). This happens when particle deformations in the system are able to finally relax (Fig.~S7). The fact that the stress continuously propagates through the system is qualitatively illustrated in the stress maps reported in Fig.~S8 for different times and can be better visualized in the Supplementary Movie. 
This phenomenology is completely absent in HZD (Fig.~S9), where stress is absent and particles overlaps are spread out, so that no persistent motion is observed.

\begin{figure}[h!]
\hspace{-0.04cm}
\centering
\includegraphics[width=1.0\textwidth]{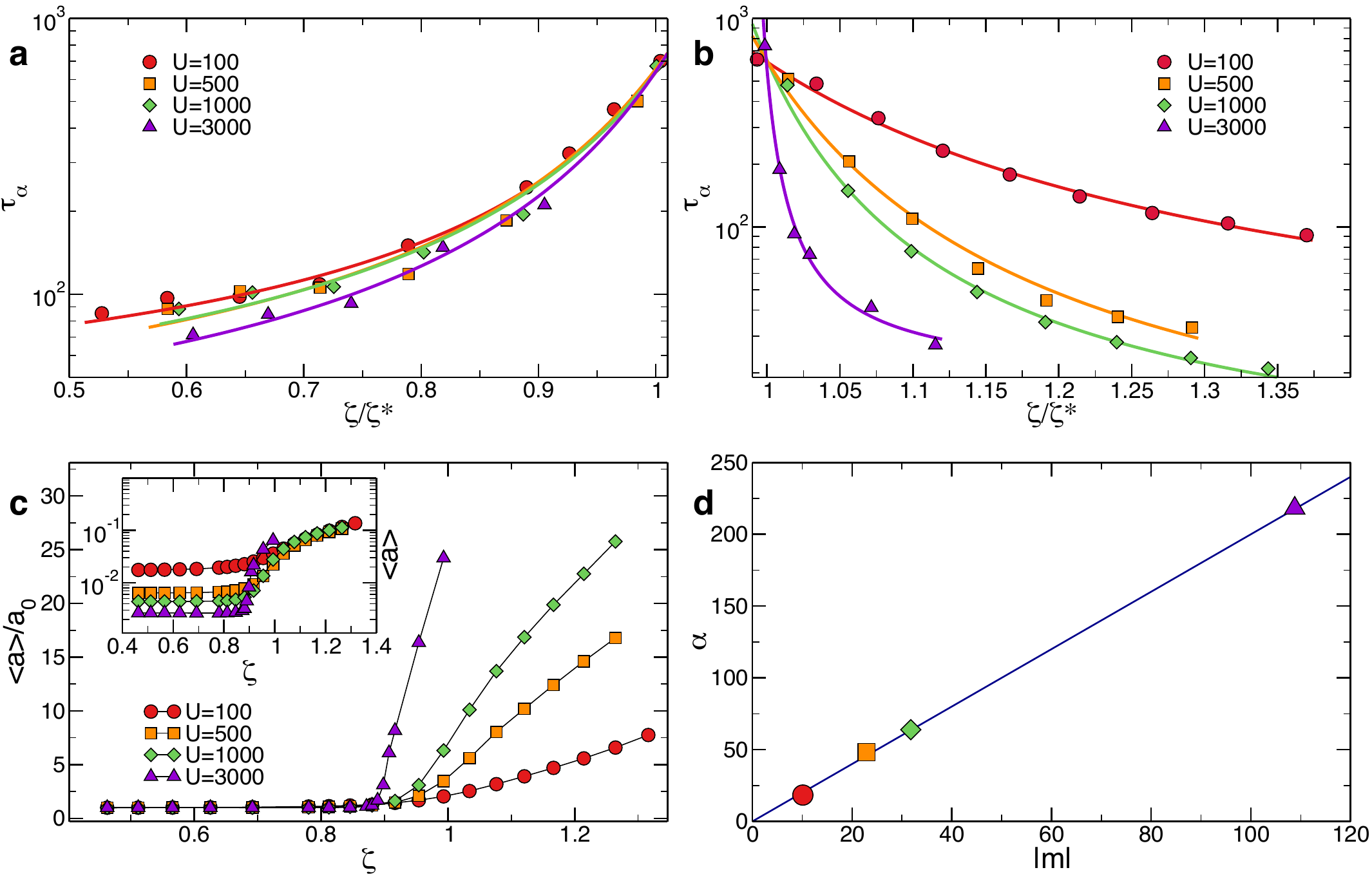}
\centering
\caption{{\bf Softness-dependent fragility}: Modified Angell's plots of the relaxation time $\tau_{\alpha}$ (a) below and (b) above melting. For each $U$ we define $\zeta^*$  as the packing fraction where $\tau_{\alpha}\simeq 650$ in reduced units; (c) average asphericity $\langle a \rangle$, normalized by its low density value $a_0=\langle a (\zeta=0.463) \rangle$, as a function of $\zeta$ for all investigated $U$. The inset reports the same data without the low-density normalization; (d) absolute value of the fragility $m=(d \tau_{\alpha}/d\zeta)|_{\zeta=\zeta^*}$ as a function of the asphericity variation $\alpha=[(d \langle a\rangle/d\zeta)\arrowvert_{\zeta=\zeta^*}]/a_0$.}
\label{fig:Panel4}
\end{figure}

The results derived so far show that particle deformations give rise to large dynamical heterogeneities which result in an intermittent collective motion of portions of particles that move ballistically.  We then address how deformation, and hence particle softness,  influences the behavior of $\tau_{\alpha}$ as a function of $\zeta$. To this aim, we report a modified Angell's plot~\cite{angell1995formation} of the relaxation time in Fig.~\ref{fig:Panel4}, obtained as discussed in Methods, both above and below $\zeta_R$. 
Interestingly, we find that for $\zeta < \zeta_R$, $\tau_{\alpha}$ is almost independent of softness (Fig.~\ref{fig:Panel4}(a)), while for $\zeta > \zeta_R$ a striking variation of the relaxation time with $\zeta$ is found (Fig.~\ref{fig:Panel4} (b)). In this regime, a transition from fragile behavior at large stiffness to (quasi)-strong behavior for soft rings is observed, the latter being characterized by an almost Arrhenius dependence of $\tau_{\alpha}$ on $\zeta/\zeta^*$.
Since this transition only occurs above melting, when particles are strongly compressed, a clear connection must exist between softness (in terms of the single-particle elastic properties) and fragility, the latter here defined as $m=(d \ln \tau_{\alpha}/d\zeta)|_{\zeta=\zeta^*}$ in analogy with that of standard glassy systems.  In our model, softness is intimately related to particle deformation. Hence, we test whether a connection between deformation and fragility also exists.
The average asphericity $\langle a\rangle$, normalized by its low-$\zeta$ value $a_0$, is shown as a function of $\zeta$ in Fig.~\ref{fig:Panel4}(c). Clearly, for $\zeta < \zeta_R$, $\langle a\rangle$ changes very little, although it varies significantly for different $U$ (inset of Fig.~\ref{fig:Panel4}(c)). However, for $\zeta \gtrsim \zeta_R$,  the variation of $\langle a\rangle$ becomes much larger for stiffer rings. We then plot  the fragility against the variation of the average asphericity $\alpha=[(d\langle a\rangle/d\zeta)|_{\zeta=\zeta^*}]/a_0$ in Fig.~\ref{fig:Panel4}(d) for  $\zeta \gtrsim \zeta_R$. A linear relation between the two quantities is found for
all investigated values of $U$, confirming that in our model a change of the single-particle internal elasticity affects the fragility of the system.

The microscopic origin of fragility in soft colloids has remained mostly elusive so far. Several experiments reported evidence of a fragility variation~\cite{mattsson2009soft,seekell2015relationship,nigro2017dynamical,nigro2018structural}  and a simple theoretical model~\cite{van2017fragility} recently showed that the osmotic regulation of compressible particles can describe a fragile-to-strong transition when the nominal packing fraction $\zeta$ is used. In Ref.~\cite{van2017fragility} this was shown to derive from the strong non-linear relationship between $\zeta$ and $\phi$, which always gives rise to a fragile behavior when the Angell's plot is represented as a function of $\phi$. Hence the observed fragility variation was interpreted as being only an apparent one. However, such a model~\cite{van2017fragility}  as well as subsequent simulations\cite{higler2018apparent} did not take into account particle deformation. Instead,  in the present study  the microscopic connection between deformation and fragility, highlighted in Fig.~\ref{fig:Panel4}(d), induces a variation of fragility at high densities,
also when the modified Angell's plot is shown as a function of $\phi$, as reported in Fig.~S10.

It is worth noting that Refs.~\cite{mattsson2009soft,nigro2017dynamical,nigro2018structural} deal with interpenetrated network microgels where charge effects such as ion-induced deswelling may be relevant, a situation far more complex than what can be described by our simple model. However, our work sheds light on recent experiments~\cite{philippe2018glass} for standard microgels and charged colloids which reported no fragility variation upon increasing $\zeta$ in the same region where we also do not detect it, i.e. for $\zeta < \zeta_R$. 
We find that it is only for $\zeta > \zeta_R$, in a regime where the rings are clearly compressed and deformed, that this variation should be detected. It may be therefore necessary that highly dense conditions, where particles are in direct contact to each other, are probed in order to observe this behaviour. 
In this regime Ref.~\cite{philippe2018glass} also reports an increase of the shape parameter $\beta$, in agreement with the present  findings, but the occurrence of solid-like behavior preempts the investigation of the system at even higher densities.  In this respect, softness is truly a valuable parameter, because highly dense states could be in principle accessed ``in equilibrium'' for much softer systems. 
To this aim, ultrasoft microgels, i.e. microgels in the absence or with very few crosslinkers~\cite{frisken2003,Bachman_2015, Virtanen_2016}, as well as hollow microgels~\cite{Scotti_2018}, for which an empty core is surrounded by a fluffy polymeric corona, may be suitable candidates to verify the present results. Our EPR model was indeed inspired from these systems, offering a simple 2D schematization of these particles, but still retaining the minimal ingredients to describe complex phenomena at high densities, such as particle deswelling and faceting.  Its 3D extension will thus be a natural perspective of this work.

\subsection*{Methods} 

\noindent{\bf Model and Simulations:} We model 2D soft particles as  polymer rings interacting with the classical bead-spring model~\cite{grest1986molecular}  with an additional internal elasticity. Each ring is composed of $N_m$ monomers of diameter $\sigma_{m}$. Two bonded monomers at distance $r$ interact through the sum of a WCA potential 
\begin{eqnarray}
V_{\rm WCA}(r) = 
\begin{cases}
4\epsilon\left[\left(\frac{\sigma_m}{r}\right)^{12} - \left(\frac{\sigma_m}{r}\right)^6 \right] + \epsilon & {\rm if} \quad r \leq 2^\frac{1}{6}\sigma_m \\[0.5em]
0 & {\rm if} \quad r > 2^\frac{1}{6}\sigma_m
\end{cases}
\label{eq:WCA}
\end{eqnarray}
\noindent
and a FENE potential 
\begin{eqnarray}
\label{eq:V_FENE}
V_{\rm FENE}(r)=-\epsilon \, k_{F}R_F^2 \ln\left[1-\left(\frac{r}{R_F\sigma_m}\right)^2\right] \quad  {\rm if} \quad r < R_F\sigma_m
\end{eqnarray}
\noindent
with $k_F=15$ the spring constant, $R_{F}=1.5$ the maximum extension of the bond and $\epsilon$ the unit of energy. Non-bonded monomers interact with  $V_{\rm WCA}(r)$ only. 
To modulate the internal elasticity of the ring we add a Hertzian potential acting between each monomer and the center of mass of the ring, as
\begin{equation}
V_H(r)=U (1-2r/\sigma_H)^{5/2}\Theta(1-2r/\sigma_H)
\label{eq:hertzian}
\end{equation}
where $U$ is the Hertzian strength in units of energy $\epsilon$ and $\sigma_H$ is the distance at rest of each monomer from the centre of mass of the ring when the polymer ring is perfectly circular. We also define the diameter of the circle inscribing the polymer ring as $\sigma_{ring}=\sigma_{H}+\sigma_{m}$. 
The addition of the internal elasticity on one hand avoids the flattening of the rings upon increasing packing fraction and on the other hand provides a tunable softness to the ring. The smaller the Hertzian strength, the larger is the ability of the ring to deform (i.e. higher softness), while with increasing $U$ the ring behaviour will tend to that of a hard disk. Thus crucially, the model takes into account particle deformation, which becomes more and more relevant at high densities (see SI), with the inner Hertzian field playing the role of an effective many-body term. 

In our study we investigate the static and dynamic properties of elastic rings for several values of $U=100, 500, 1000$, $3000$. We mainly show results for $N=1000$ elastic rings with $N_{m}=10$ monomers for which $\sigma_H=3.107\sigma_m$. However, we note that we also tested the cases of $N_{m}=5$ and $N_{m}=20$ for which we found similar results. In addition we also examined a larger system composed of $N=10000$ rings and  $N_{m}=10$ for selected high packing fractions (see Fig.~S11). Due to the high propensity of the system to crystallize we use a size polydispersity of $12\%$ both for $\sigma_m$ and for $\sigma_H$  according to a log-normal distribution.

We perform Langevin dynamics simulations at constant temperature with $k_BT=1$. We use as unit of length the average ring diameter $\langle\sigma_{ring}\rangle$ at low dilution and as unit of time $t_0=\langle\sigma_{ring}\rangle \sqrt{m_{ring}/\epsilon}$ where $m_{ring}=m\cdot N_{m}$ and $m$ is the monomer unit mass. A velocity Verlet integrator is used to integrate the equations of motion with a time step $dt=10^{-3}$. We follow Ref. ~\cite{russo2009reversible} to model Brownian diffusion by defining the probability $p$ that a particle undergoes a random collision every $Y$ time-steps for each particle. By tuning $p$ it is possible to obtain the desired free particle diffusion coefficient $D_0=(k_BTY dt/m)(1/p -1/2)$. We fix $D_0=0.008$ but we also checked other values ranging from $D_0=0.0015$ to $D_0=0.08$, finding that there is no influence of the microscopic dynamics on the long-time behavior (see Fig.~S12). The packing fraction of the rings is defined as $\zeta=\sum_{i=1}^N \pi\sigma^{2}_{i, ring}/4L^2$, where $\pi\sigma^{2}_{i, ring}/4$ is the area of the $i$-th ring at low dilution. 

\noindent \textbf{Rings Deformation:} To evaluate how polymer rings deform, we calculate the gyration tensor from which we extract the radius of gyration $R_g=[1/(N_m) \sum_{i=1}^{N_m} (\vec{r}_{i}-\vec{r}_{CM})^2]^{1/2}$, where $\vec{r}_{CM}$ is the centre of mass of the ring. We also calculate the asphericity parameter as $a=(\lambda_2-\lambda_1)^2/(\lambda_1+\lambda_2)^2$ where $\lambda_1$ and $\lambda_2$ are the eigenvalues of the gyration tensor~\cite{rudnick1986aspherity}.
In addition we also estimate the effective packing fraction $\phi$ occupied by the rings by calculating the average area $A_R$ of rings from the gyration radius of the each ring or from the area of the ellipse having as semi-axis the eigenvectors extracted from the gyration tensor. As both approaches yield similar results, we use the average between the two.  Once the area is calculated, the effective packing fraction is obtained as $\phi=\zeta A_R/A_{R_0}$, where $A_{R_0}$ is the average value of the area of a ring at low dilution.

\noindent \textbf{Rings Dynamics}: We have quantified the dynamics of the rings by evaluating the mean-squared displacement $\langle r^2(t)\rangle=(1/N)\langle \sum_{i=1}^N (\vec{r}^{\ i}_{CM}(t)-\vec{r}^{\ i}_{CM}(t=0))^2\rangle$ (where $\langle \dots \rangle$ denotes an ensemble average) and the self-intermediate scattering function $F_s(q^*,t)=(1/N)\langle \sum_{i=1}^N e^{i\vec{q}^*\cdot (\vec{r}^{\; i}_{CM}(t)-\vec{r}^{\; i}_{CM}(t=0))}\rangle$.   Both observables are calculated using the positions of the centers of mass of the rings. 
It is well established that the long-time decay of $F_s(q^*,t)$ can be described by a generalized exponential decay $F_s(q^*,t)\sim \exp[-(t/\tau_{\alpha})^\beta]$ modulated by a ``shape parameter'' $\beta$. To extract the value of $\beta$ and $\tau_{\alpha}$, we approximate the whole $F_s(q^*,t)$ as the sum of two exponentials:
\begin{equation}
F_s(q^*,t)=C\exp(-t/\tau_{0})+(1-C)\exp(-(t/\tau_{\alpha})^\beta),
\label{eq:fitbeta}
\end{equation} where the first one is a simple exponential which accounts for the short-time decay controlled by $\tau_0$ and the second one provides a description of the long-time decay, with $C$ a constant varying between 0 and 1. When the dynamics becomes very fast at high $\zeta$, only the second exponential is retained in Eq.~\ref{eq:fitbeta}, being a single compressed exponential able to interpolate the whole curve.

\noindent \textbf{Analysis of fast rings}: At high $\zeta$ we have divided the total simulation time into windows of duration $\Delta t$ of the order of $\tau_{\alpha}$. In particular we fix $\Delta t=7.88$ in reduced units and we select the $10\%$ fastest particles in each window. We then calculate the MSD only for these particles, starting from the interval where they were selected. The dynamics of fastest rings, quantified by their MSD averaged over all considered windows, is faster than that of all rings, i.e. it can be described by a larger $\gamma$ exponent of the superdiffusive behavior in the MSD that is found to be close to 1.9. However, at larger times, they always retain diffusive behavior. 
It is important to note that, despite the large polydispersity of the system, fastest particles are evenly distributed among all particle sizes.
Analyzing different time windows, we find that for the considered state point ($U=1000$ and $\zeta=1.264$) the super-diffusive exponent is exactly equal to $2$, signalling purely ballistic dynamics, for about 25\% of the considered time intervals. This ratio strongly depends on the chosen state point and on the number of considered fastest rings. Indeed, for each interval, we can always define a subset of  fastest rings undergoing purely ballistic dynamics; their number increases with increasing $\zeta$ and increasing $U$, thus explaining the behavior of the average value of $\gamma$, reported in Fig.~S4, and also of the shape parameter $\beta$, shown in Fig.~\ref{fig:Panel1}(d).

\noindent \textbf{Modified Angell's plot:} In standard glass-formers an Angell's plot reports the variation of the viscosity or relaxation time of the system in a logarithmic scale versus the control parameter driving the glass transition, for example the packing fraction in the case of colloidal suspensions, appropriately rescaled by its glass transition value\cite{angell1995formation,debenedetti2001supercooled,van2017fragility}. In our system, we do not find a glass transition, but the system undergoes a reentrant melting. We thus use the value of $\zeta_R$ as rescaling packing fraction and visualize the dependence on the relaxation time on $\zeta$, approaching this value both from above and from below. However, rings with different softness $U$ have different relaxation times at $\zeta_R$, so that it is more convenient to use a packing fraction $\zeta^*$, slightly different but close to $\zeta_R$ and dependent on $U$, for whic a common value of $\tau_{\alpha}$ is found. This allows us to scale altogether the data for different values of $U$. In the text, we report results where we have chosen the common value $\tau_{\alpha}\simeq 650$ in reduced units, but we have verified that different choices provide qualitative similar results.

\subsection*{Acknowledgments}
We thank F. Camerin, L. Cipelletti, C. Maggi, A. Ninarello and D. Truzzolillo for useful discussions and comments. 
We acknowledge support from the European Research Council (ERC Consolidator Grant 681597, MIMIC) and from ETN-COLLDENSE (H2020-MCSA-ITN-2014, Grant 642774). 

\renewcommand{\thefigure}{S\arabic{figure}}
\setcounter{figure}{0}

\section*{Supplemental Information}

\subsection{Particle deformation on increasing packing fraction}
\begin{figure}[h]
\hspace{-0.04cm}
\centering
\includegraphics[width=1.0\textwidth]{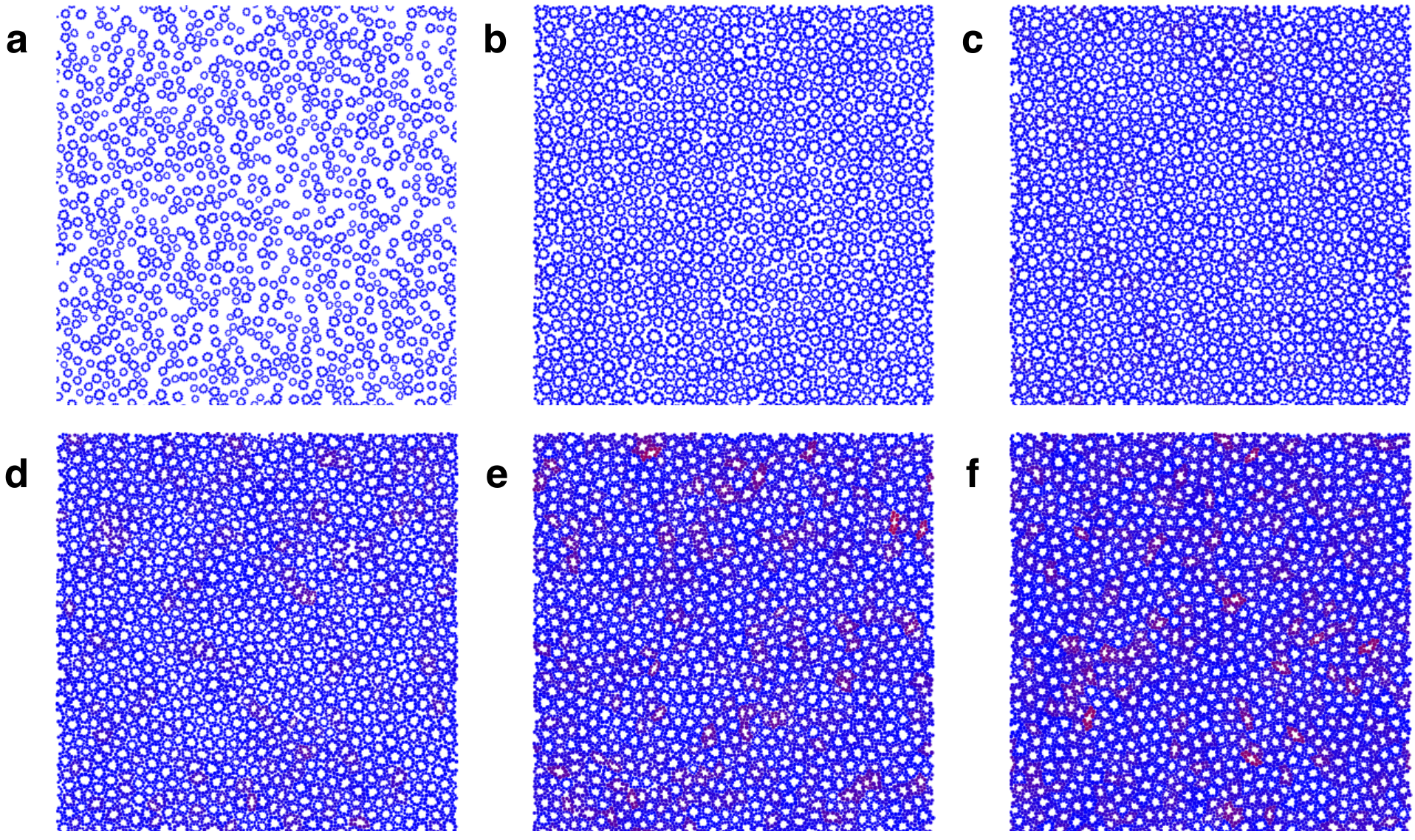}
\centering
\caption{(a)-(f) Snapshots of elastic polymer rings with $U=1,000$ at $\zeta=0.463, 0.812, 0.917, 1.034, 1.167, 1.264$. Rings are coloured from blue (not deformed) to red (highly deformed) according to their asphericity.}
\label{fig:S1}
\end{figure}

Figure~\ref{fig:S1} reports snapshots of EPRs at different packing fractions for $U=1,000$. To show the degree of deformation each EPR is coloured according to its asphericity $a$: circular rings are blue while strongly deformed rings are coloured in red. In the more compressed conditions, faceting is evident between strongly deformed rings. Movies of rings at three different packing fractions $\zeta=0.463, 0.812$ and $1.264$ are reported in the Supplementary Movie where EPRs change colours in time according to their asphericity following the colour code of Fig.~\ref{fig:S1}.
Each movie is composed of frames separated by a time of $\sim 40$ (in reduced units) for up to a total time of $800$. 

\subsection{Dynamics of Hertzian disks}
To compare results for EPRs with systems in which only interpenetration is allowed, we also perform Langevin dynamics simulations of $N=10^3$ Hertzian disks (HZDs) interacting with $V_{HZ}=U_{H}(1-r/\sigma)^{5/2}\Theta(1-r/\sigma)$, where $\sigma$ is the disk diameter and we fix $U_{H}=200$ in energy units $\epsilon$. Disks are polydisperse with the same log-normal distribution used for the EPRs with a polydispersity of $12\%$. We define a time unit $t_0=\langle\sigma\rangle\sqrt{m/\epsilon}$  where $m$ is the mass of a disk and $\langle\sigma\rangle$ is the unit of length.
\begin{figure}[h]
\hspace{-0.04cm}
\centering
\includegraphics[width=1.0\textwidth]{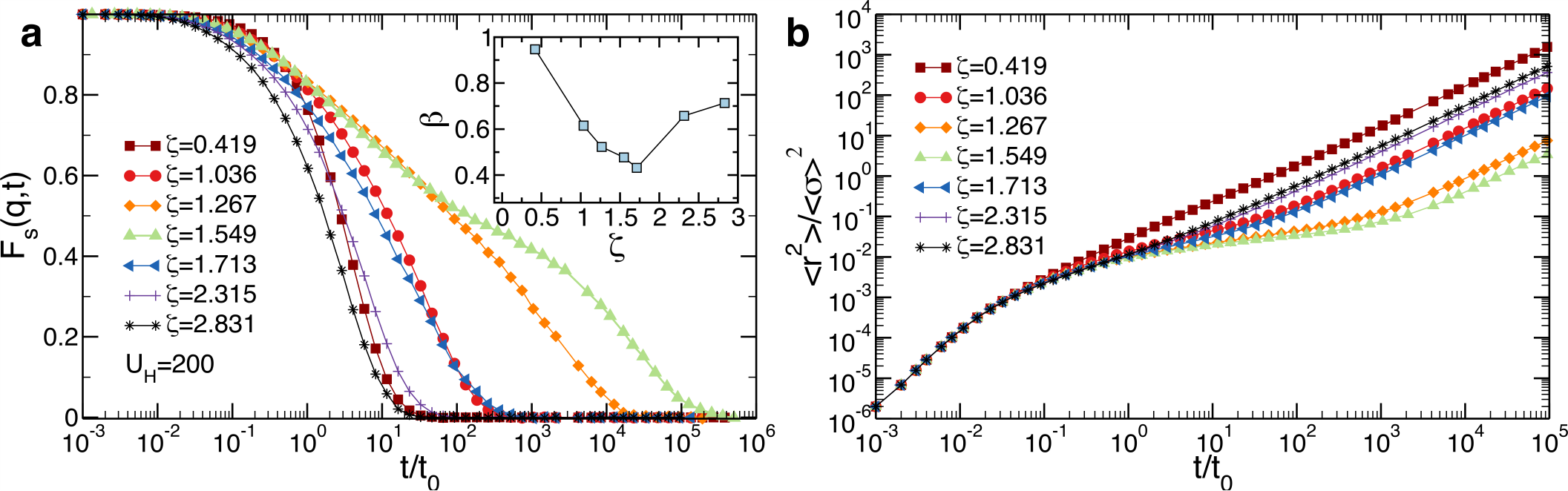}
\centering
\caption{(a) Self-intermediate scattering functions $F_{s}(q^{*},t)$ and (b) mean-squared displacements $\langle r^2(t)\rangle$ for $N=10^3$ Hertzian disks with strength $U_H=200$ at different $\zeta$. Inset of (a):  shape parameter $\beta$ of $F_{s}(q^{*},t)$ obtained via an exponential fit of the long-time decay.  }
\label{fig:S8}
\end{figure}

Figure~\ref{fig:S8} shows dynamical and structural properties of  Hertzian disks at different $\zeta$. As for EPRs, also HZDs display a reentrant transition towards a fluid at high $\zeta$. However, differently from EPRs, the $F_{s}(q^{*},t)$ do not display a compressed exponential decay as shown also in the inset of  Figure~\ref{fig:S8}(a)  where the shape parameter $\beta$ as a function of $\zeta$ is found to be always smaller than $1$.  
Similarly, the MSD reported in Fig.~\ref{fig:S8}(b) does not show a super-diffusive behavior at any studied $\zeta$.

\subsection{Dynamics of rings of different softness}

\begin{figure}[h]
\hspace{-0.04cm}
\centering
\includegraphics[width=1.0\textwidth]{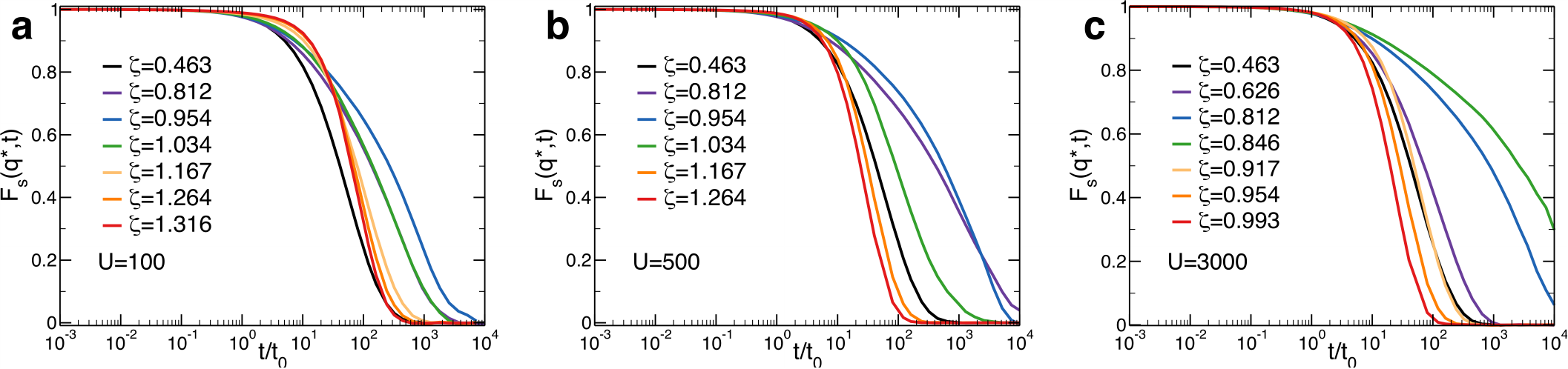}
\centering
\caption{Self-intermediate scattering functions $F_s(q^*,t)$ for EPRs with (a) $U=100$, (b) $U=500$ and (c) $U=3,000$ as a function of $\zeta$. }
\label{fig:S3}
\end{figure}

We report the self-intermediate scattering function $F_s(q^*,t)$ (Fig.~\ref{fig:S3}) of EPRs with Hertzian field of different amplitudes, i.e. $U=100; 500$ and $3,000$. In all cases we observe the same qualitative features described in the main text  for $U=1,000$:  a reentrant melting upon increasing $\zeta$ and a compressed exponential decay of $F_s(q^*,t)$. We further note that at $q=q^*$, self and collective intermediate scattering functions give similar results (not shown).

\begin{figure}[h!]
\hspace{-0.04cm}
\centering
\includegraphics[width=0.4\textwidth,clip]{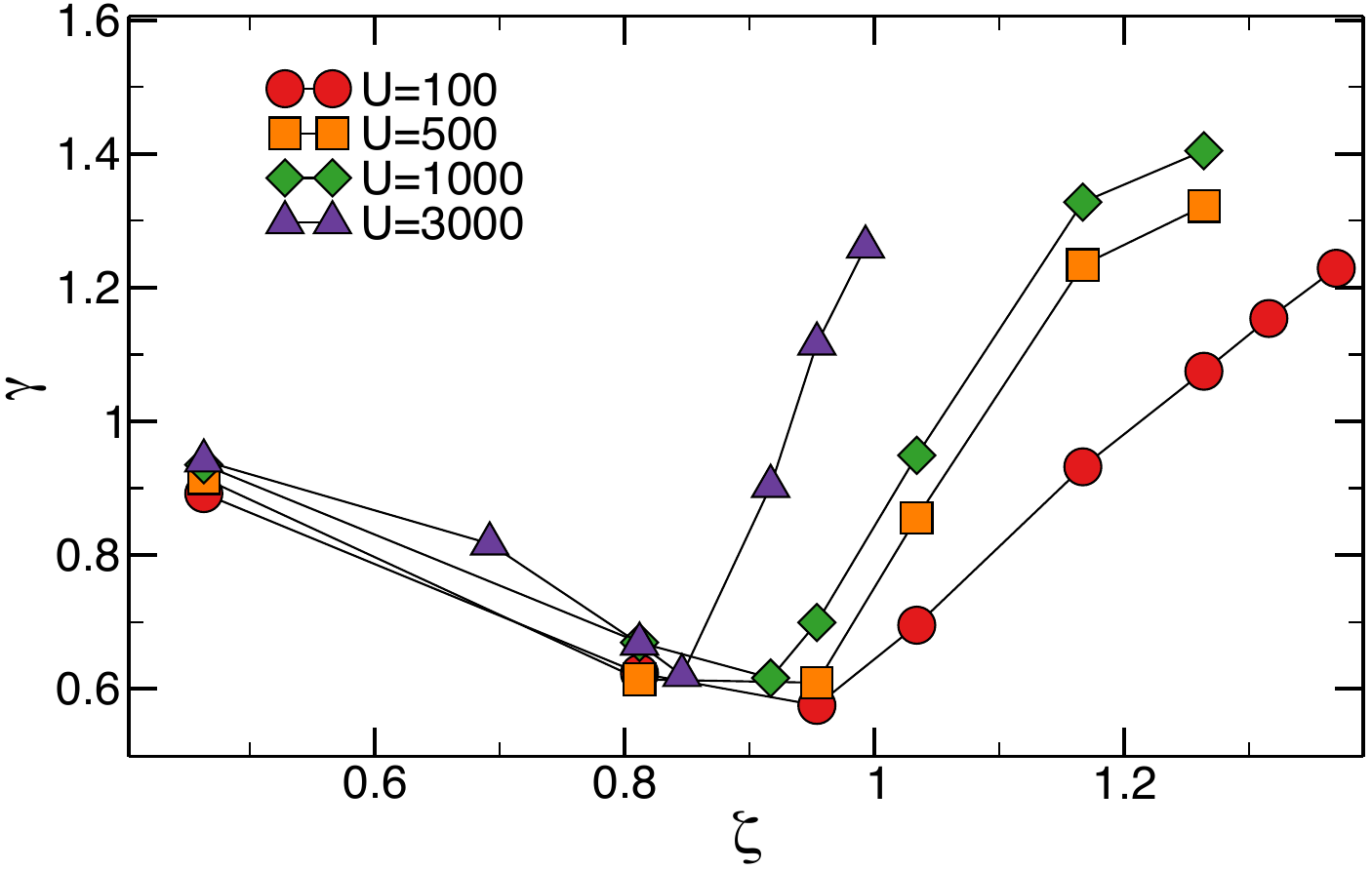}
\centering
\caption{Evolution of the $\gamma$ exponent  for EPRs with different $U$ as a function of $\zeta$. $\gamma$ is extracted from $\langle r^2(t)\rangle$ using the interpolating function $\langle r^2(t)\rangle\sim t^{\gamma}$.} 
\label{fig:S5}
\end{figure}
\newpage
Figure~\ref{fig:S5} quantifies the dependence of the $\gamma$ exponent which controls the time dependence of  $\langle r^2(t)\rangle\sim t^{\gamma}$ at time scales where the superdiffusive regime is observed for EPRs at different values of $U$. We extract $\gamma$ from a power-law fit of $\langle r^2(t)\rangle$ at intermediate times, finding a similar behavior in $\zeta$ and $U$ as that observed in the evolution of the shape parameter $\beta$, discussed in the main text. 

\subsection{Distribution of particle asphericity}
\begin{figure}[h]
\hspace{-0.04cm}
\includegraphics[width=0.5\textwidth,clip]{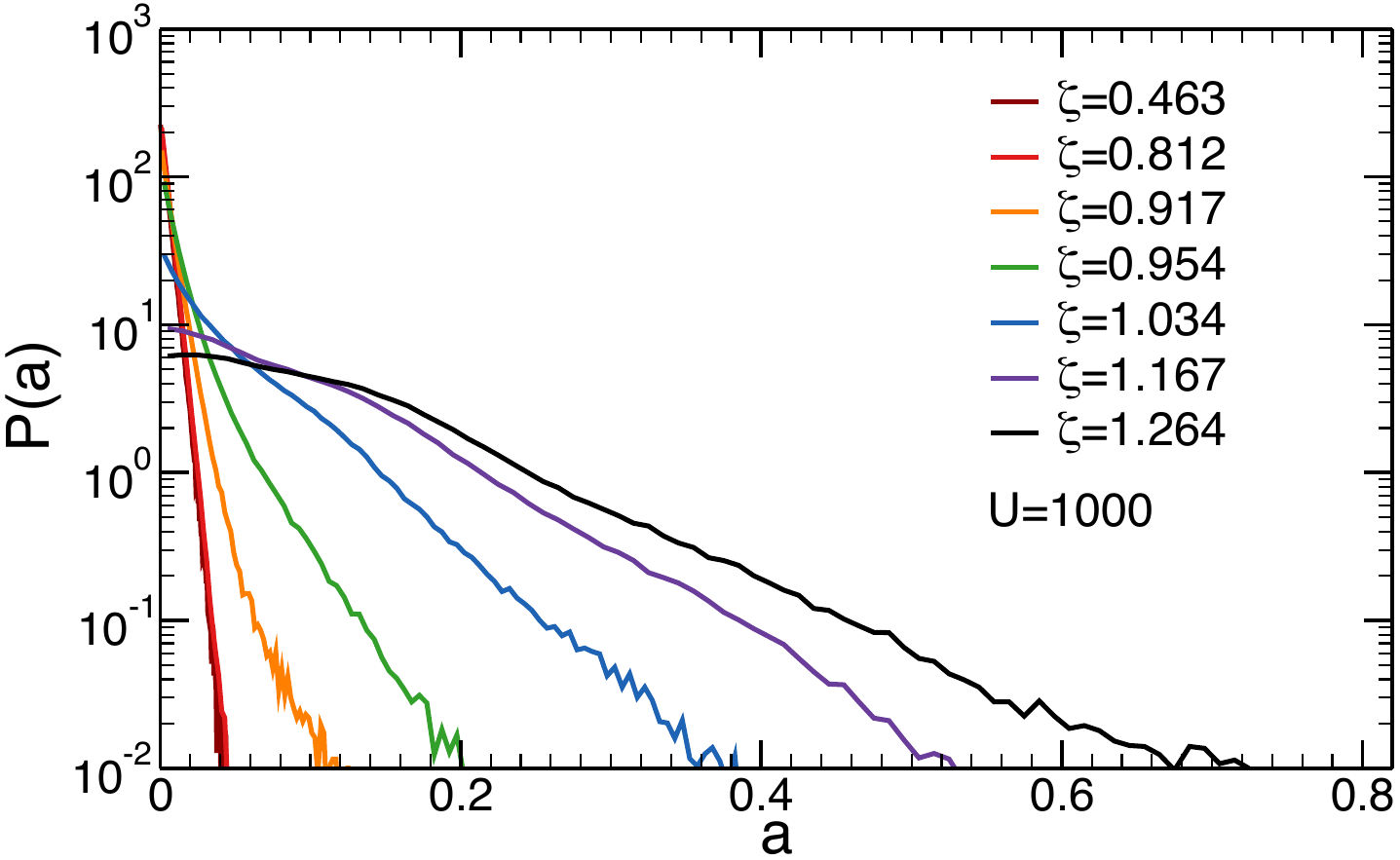}
\centering
\caption{Normalized distribution of particle asphericity $P(a)$ for EPRs with $U=1,000$ at different $\zeta$. }
\label{fig:S6}
\end{figure}
The distributions of particle asphericity $P(a)$  for EPRs are shown for $U=1,000$ as a function of $\zeta$ in Fig.~\ref{fig:S6}.  We find that the shape of $P(a)$ strongly changes upon increasing $\zeta$, as a result of two main contributions: a slowly decreasing probability of finding weakly deformed particles which becomes roughly constant at high $\zeta$ and a growing exponential tail which describes the probability of finding strongly aspherical (and hence deformed) polymer rings. 

\subsection{Intra-ring stress analysis and connection to particle deformation}
\begin{figure}[h]
\centering
\includegraphics[width=0.8\textwidth,clip]{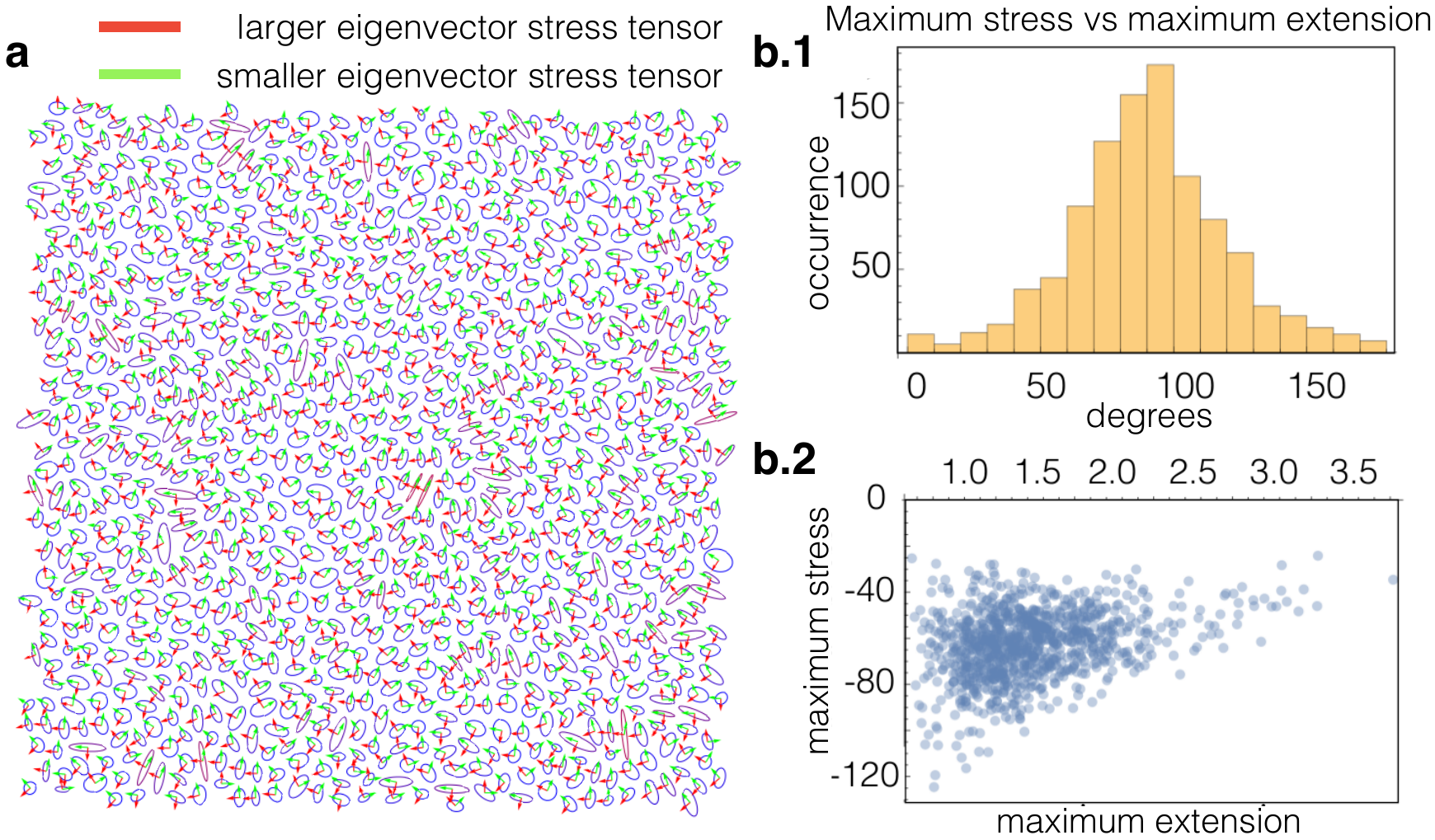}
\centering
\caption{(a) Snapshot of EPRs with $U=1,000$ and $\zeta=1.264$ drawn as ellipses built from the gyration tensor eigenvectors. The vectors drawn at the center of mass of the rings are the two normalized eigenvectors of the intra-ring stress tensor, where red/green colours indicate respectively the maximum and the minimum eigenvector; (b.1): distribution of the angles formed by the maximum eigenvector of the intra-ring stress tensor and the maximum eigenvector of the gyration tensor; (b.2) scatter plot of the maximum eigenvalue of the gyration tensor and that of the stress tensor. The latter is negative indicating the presence of a compressive stress in that direction.}
\label{fig:intra-stress}
\end{figure}
To monitor the single ring stress behaviour, we calculate the associated stress tensor~\cite{Rapaport} where, for each ring, the monomer-monomer and monomer-inner Hertzian forces are accounted. The resulting normalized eingevectors are shown for each ring in Fig.~\ref{fig:intra-stress}(a).  Clearly, a simple look at the figure makes it possible to identify some correlations between the eigenvector directions and those of the two vectors defining the semiaxes of the ellipses, being the latter by construction the eigenvectors of the gyration tensor of the ring. To better quantify this correlation, we calculate the angle formed by the maximum eigenvector of the stress and of the gyration tensor, reported in Fig.~\ref{fig:intra-stress}(b.1), finding that they are mostly orthogonal, forming angles distributed around 90$^{\circ}$. In Fig.~\ref{fig:intra-stress}(b.2) we also show that the eigenvalues of the stress and of the gyration tensor are correlated, so that a larger stress corresponds to a larger deformation (and asymmetry). The negative values of the stress indicate that this is of compressive nature in the maximum direction. This analysis clearly points out that deformation and stress are linked and confirms that the analysis on deformations presented in the main text is well-defined.

\subsection{Stress propagation in the system}
\begin{figure}[h]
\centering
\includegraphics[width=0.5\textwidth,clip]{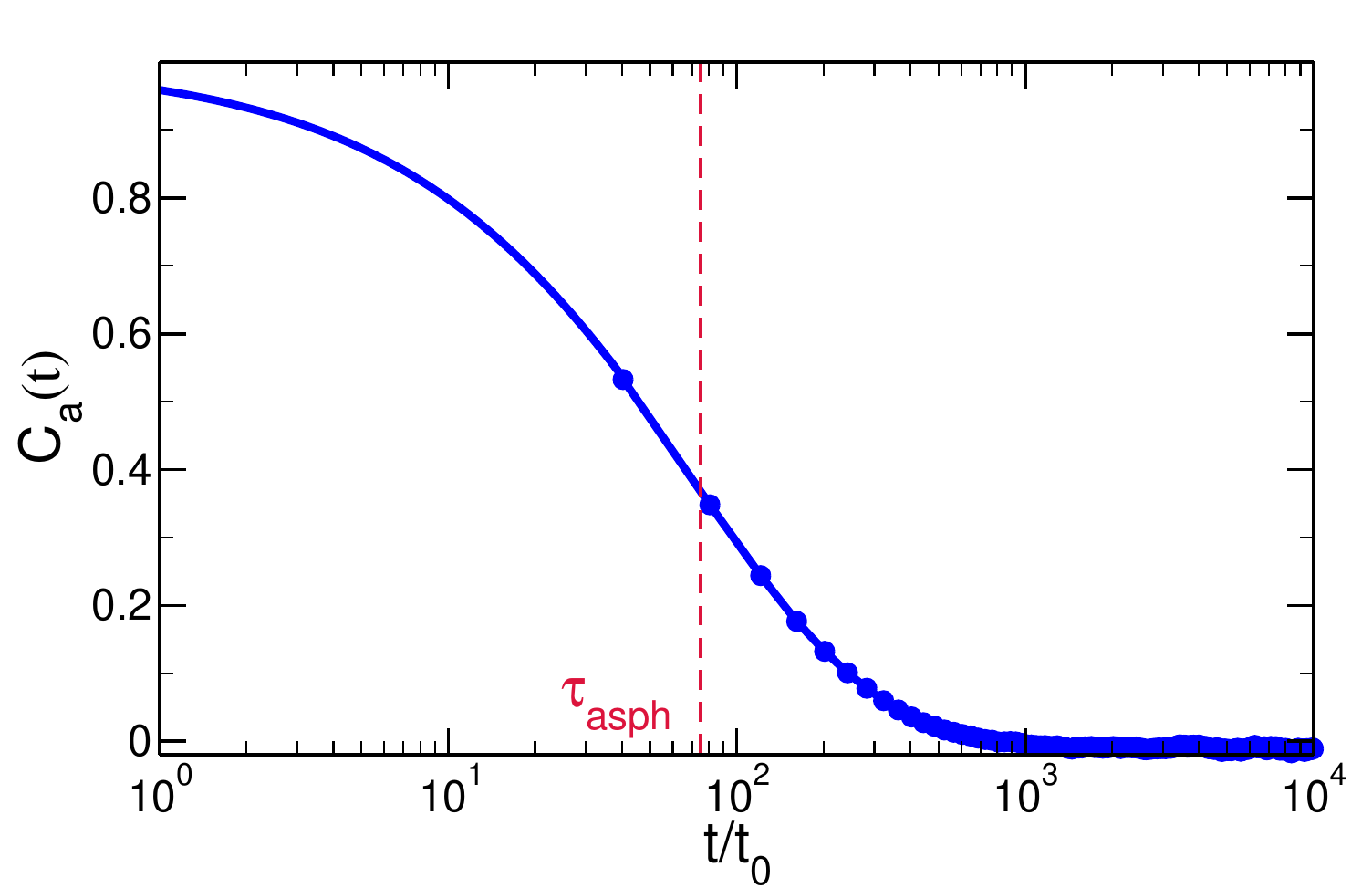}
\centering
\caption{Autocorrelation function of the asphericity for rings with $U=1,000$ at $\zeta=1.264$. The red dashed line represents the value $1/e$ at which we estimate the relaxation time of the function.}
\label{fig:Casph}
\end{figure}

To get more insight into the mechanism of stress propagation within the system we calculate the asphericity autocorrelation function $C_a(t)= \sum_{i=1}^{N} a_i(t)a_i(0)/ \sum_{i=1}^{N} a^2_i(0)$  for  $U=1,000$ and $\zeta=1.264$, shown in Fig.~\ref{fig:Casph}. 
\begin{figure}[h]
\centering
\includegraphics[width=1.0\textwidth,clip]{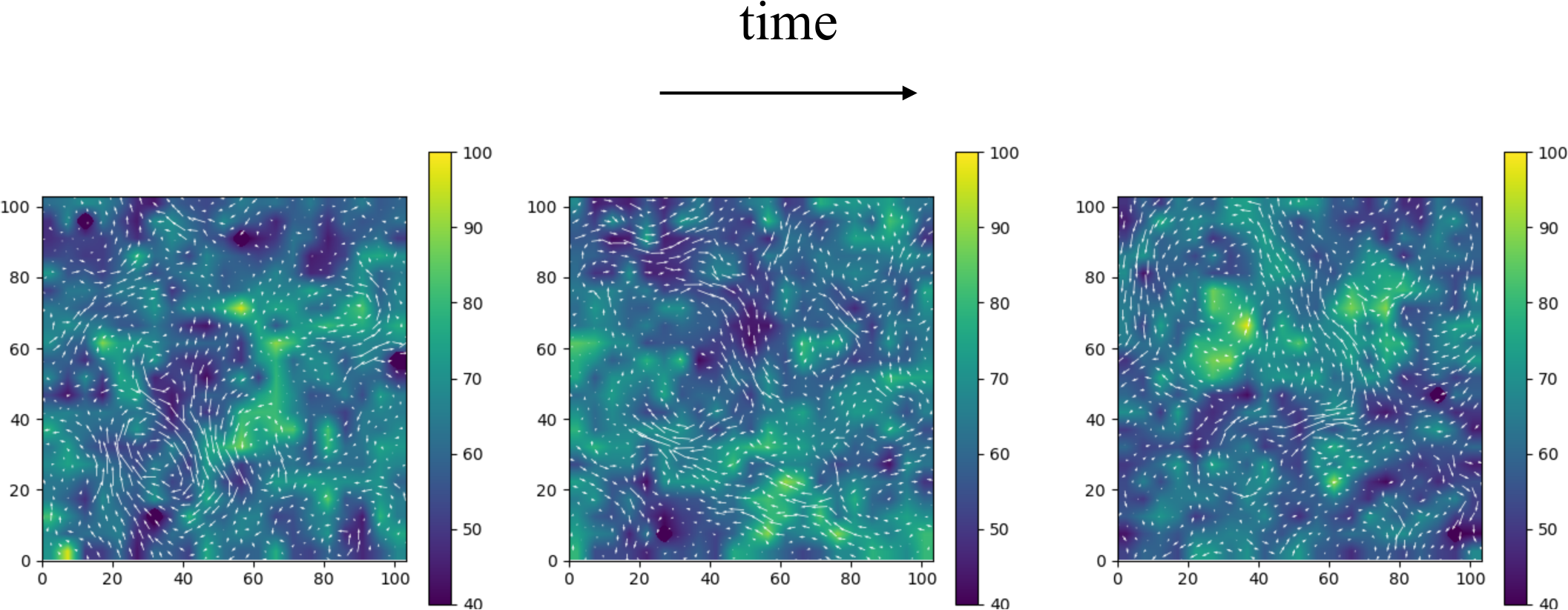}
\centering
\caption{Contour plots of the magnitude of the intra-stress largest eigenvalue averaged over the time window $\Delta t=7.88$ for rings with $U=1,000$ and $\zeta=1.264$. White arrows are particle displacements over the same time window.
The plots are separated in a time value roughly corresponding to the relaxation time of $C_a(t)$.}
\label{fig:StressMaps}
\end{figure}

An interesting result is that the relaxation time $\tau_{asph}$ of $C_a(t)$ corresponds to the upper time limit of the superdiffusive regime at the considered state point, which confirms that stress propagation and superdiffusion (and consequently the compressed exponential) are correlated phenomena. In addition, the existence of  portions of particles that move ballistically is in line with mean-field models of stress propagation in elasto-plastic materials~\cite{ferrero2014relaxation}, where ballistic motion is activated by a stress release in the medium.
Stress transmission on the time scales of the superdiffusive regime allows to decorrelate large spatial stress correlations as shown in Fig.~\ref{fig:StressMaps}.

\subsection{Fastest particles analysis for Hertzian disks}
\begin{figure}[h!]
\hspace{-0.04cm}
\centering
\includegraphics[width=0.4\textwidth,clip]{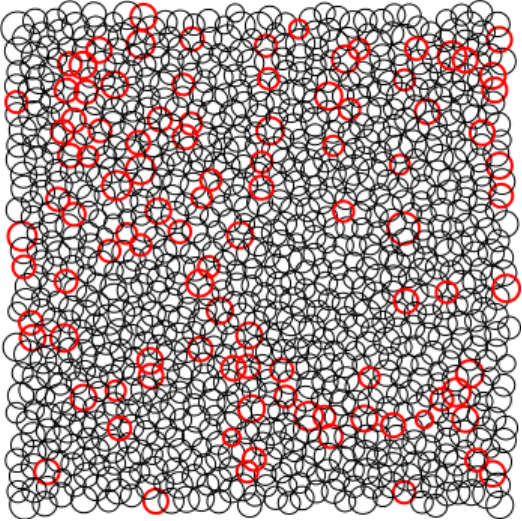}
\centering
\caption{Snapshots of Hertzian disks at $\zeta=2.831$. Red particles represent the $10\%$ fastest particles in a time window comparable to the relaxation time of the system.}
\label{fig:S9}
\end{figure}

Although Hertzian disks do not show a superdiffusive regime in the MSD, we repeat the analysis for the fastest particles done for EPRs as described in the main text. We then select time windows of length comparable to $\tau_{\alpha}$ at the state point $\zeta=2.831$, confirming that we never observe a superdiffusive or ballistic behavior of the MSD, even for a small fraction of the particles.  In addition, by highlighting fast particles in the snapshot reported in  Fig.~\ref{fig:S9}, it is clear that they are homogeneously distributed within the whole system, contrarily to what found for EPRs in Fig.~3 of the main text.

\subsection{Modified Angell's plot as a function of effective packing fraction}
\begin{figure}[h!]
\begin{center}
\includegraphics[width=0.5\textwidth,clip]{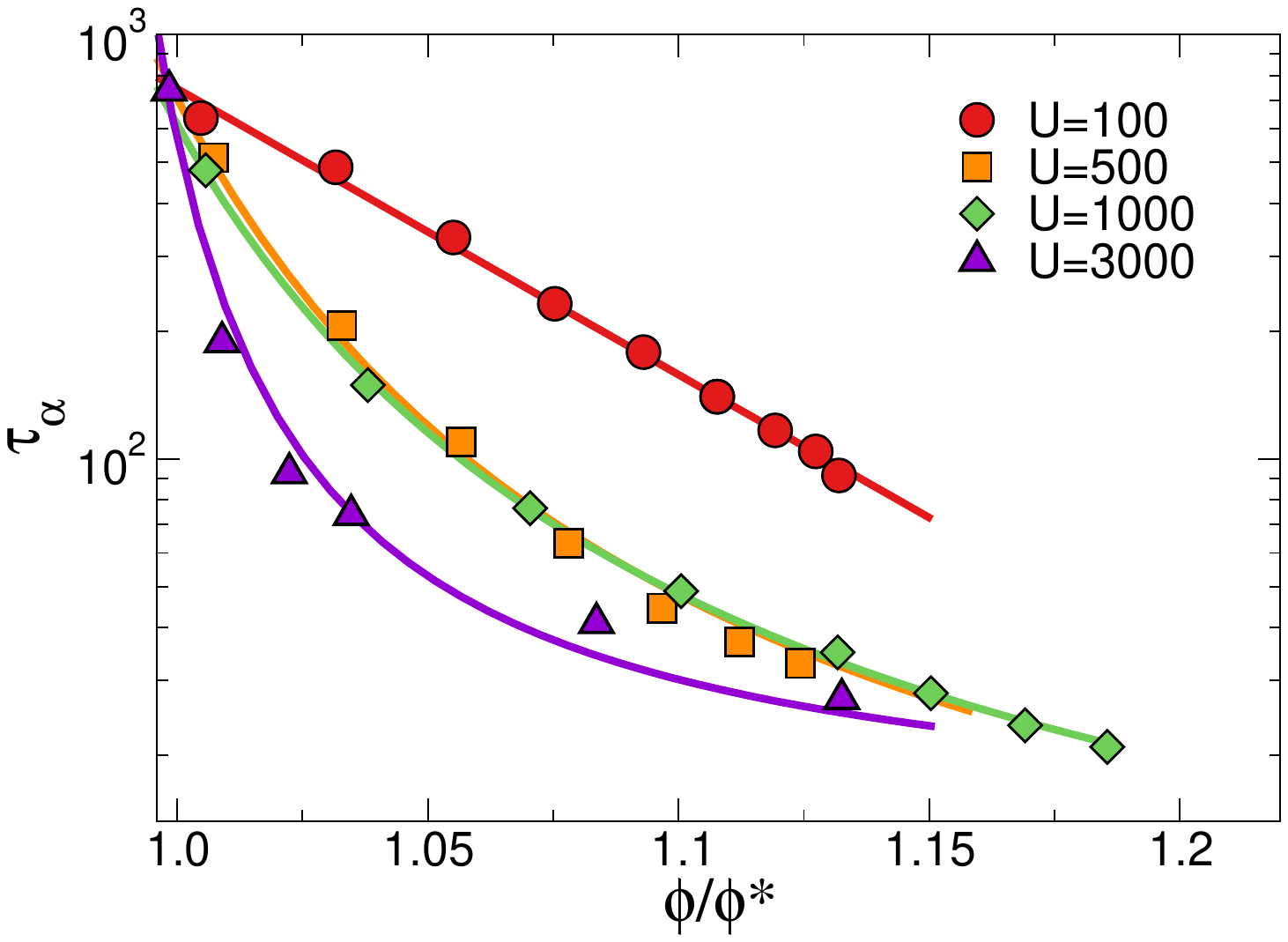}
\centering
\caption{Modified Angell's plot of the relaxation time versus effective packing fraction $\phi$ at high densities. For each $U$ we define $\phi^*$, in analogy with $\zeta^*$, as the effective packing fraction where $\tau_{\alpha}\simeq 650$ in reduced units. Solid lines are guides to the eye.}
\label{fig:Snew}
\end{center}
\end{figure}

\noindent Recent theoretical and numerical studies, based on models of soft particles that undergo osmotic deswelling only, argued that the change in fragility as a function of particle elasticity is only apparent\cite{van2017fragility, higler2018apparent}. This was based on the fact that, while the Angell's plot as a function of the nominal packing fraction $\zeta$ does show a dependence on elasticity, all the data collapse together when the effective packing fraction $\phi$ is used, instead of $\zeta$. We here show the same analysis for the EPRs combining the data in Fig.~3(a) and Fig.~4(b) of the main text. The modified Angell's plot is shown as a function of $\phi/\phi^*$ in Fig.~\ref{fig:Snew}, where $\phi^*$ is defined in the same way as $\zeta^*$, i.e. where $\tau_{\alpha}\simeq 650$ in reduced units. 

This result clearly shows that the variation of fragility as a function of softness in our system is true and not apparent, and hence a true relation between particle elasticity and  fragility exists.

\subsection{Size effects}
\begin{figure}[h!]
\hspace{-0.04cm}
\centering
\includegraphics[width=1.0\textwidth,clip]{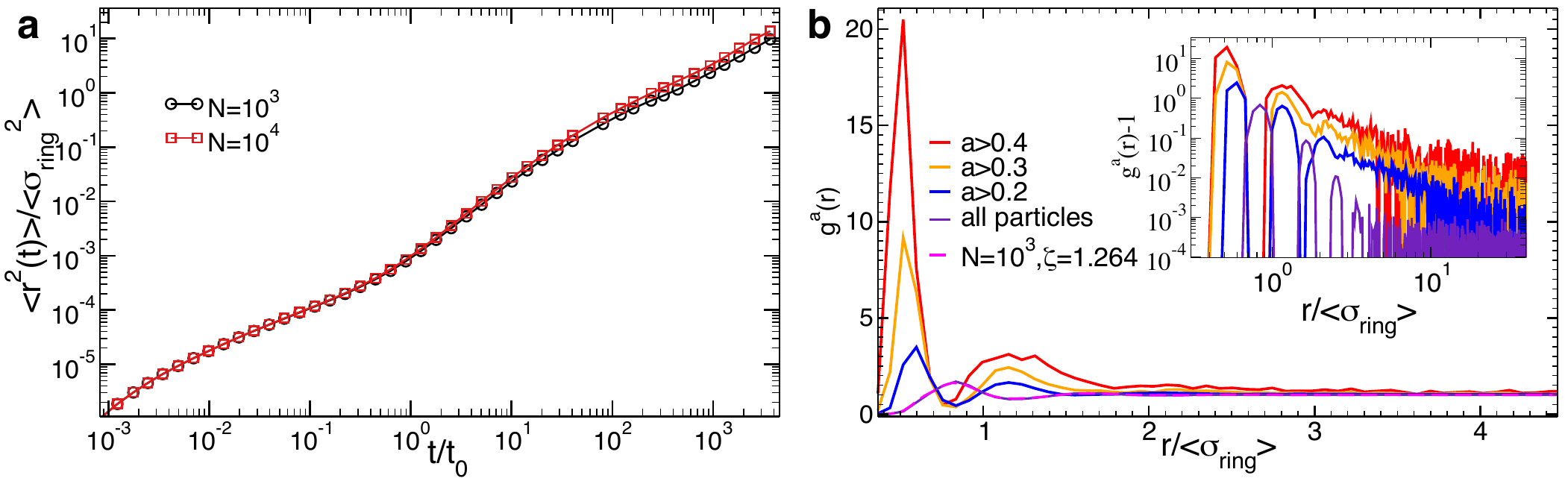}
\centering
\caption{(a) Mean-squared displacement for $U=1,000$ and $\zeta=1.264$ for two system sizes: $N=10^4$ and $N=10^3$ rings; (b) radial distribution function $g^{a}(r)$ for particles with asphericity $a> 0.2,0.3,0.4$ for the large system with $N=10^4$ and $\zeta=1.264$. Inset: $g^{a}(r)-1$ is shown to highlight the growth of a correlation length among aspherical particles.}
\label{fig:S7}
\end{figure}

The study of the dynamics in 2D systems should be taken with caution due to the presence of Mermin-Wagner long-range fluctuations that was shown to affect the glass transition of hard disks~\cite{flenner2015fundamental}. However, in the present work, these should not affect qualitatively the compressed nature of the exponential relaxation or the intermediate superdiffusive regime. To verify whether this is the case, we perform additional simulations of $N=10^4$ disks, i.e. one order of magnitude larger than the system discussed in the main text, at $\zeta=1.264$. The comparison between the MSD at different sizes is shown in Fig. \ref{fig:S7}(a). We find that the system size does not change the extension of the superdiffusive regime or the value of the exponent $\gamma$. Instead, we find a slightly larger diffusion coefficient in agreement with Ref.~\cite{flenner2015fundamental}.  We also confirm that the compressed exponential relaxation for the self-intermediate scattering function is also found.
The larger system also allows us to better investigate the structural correlations between aspherical particles via the $g^{a}(r)$ as shown in Fig. \ref{fig:S7}(b), finding that the more particles are aspherical, the more they are structurally correlated. The growth of a correlation length can be observed from the exponential decay of $g^{a}(r)-1$ in the inset of Fig.~\ref{fig:S7}(b).

\subsection{Influence of the microscopic dynamics on the superdiffusive regime}

\begin{figure}[h!]
\hspace{-0.08cm}
\centering
\includegraphics[width=0.5\textwidth,clip]{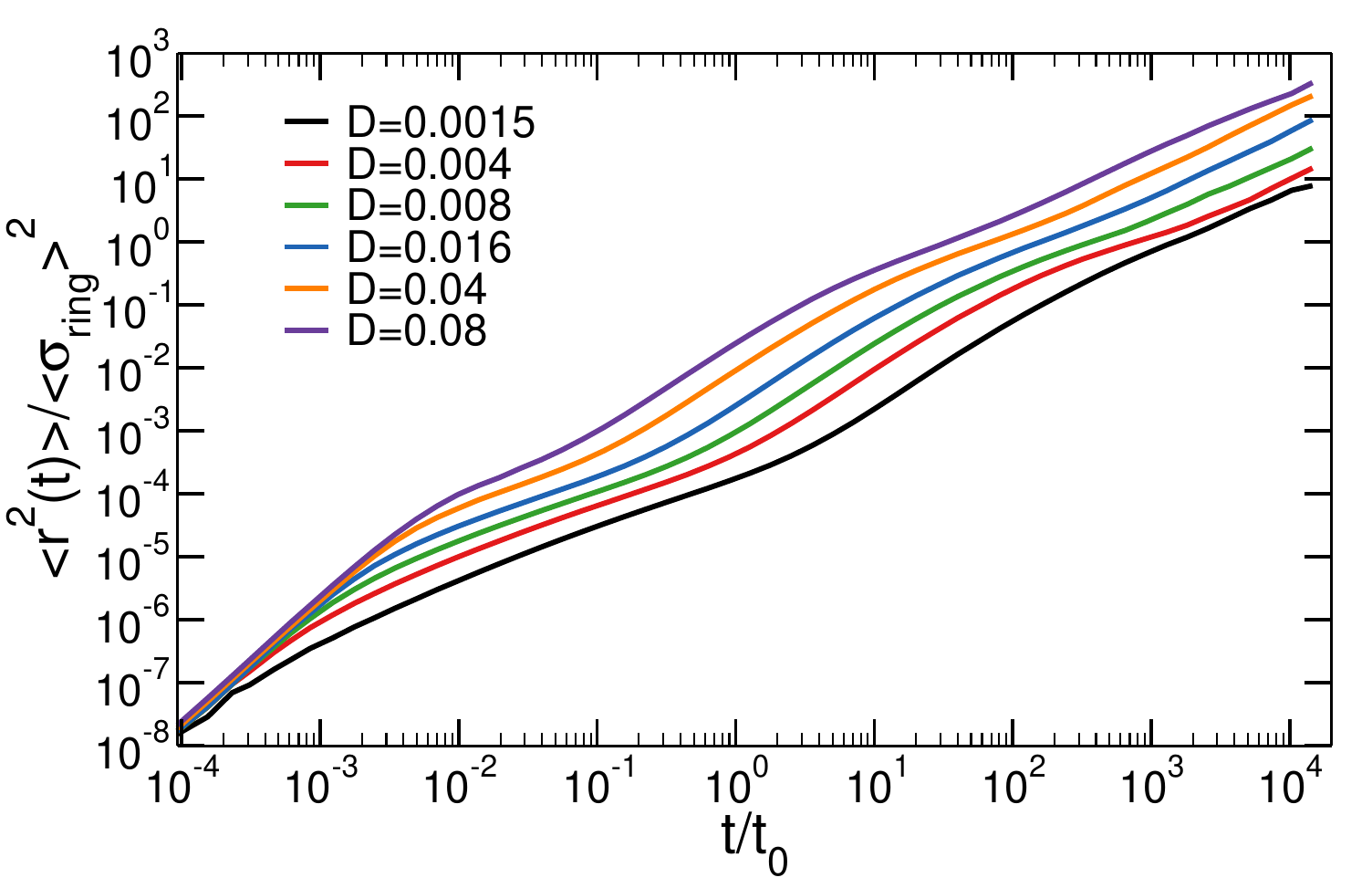}
\centering
\caption{Mean square displacement of EPR center of mass at $\zeta=1.264$ for rings with $U=1,000$ for different values of the free particle diffusion coefficient $D_0$.}
\label{fig:D0}
\end{figure}

To check whether the microscopic dynamics has an influence on the superdiffusive behavior observed in the MSD, we change the free particle diffusion coefficient $D_0$ by changing the parameter $Y$ (see Methods) and perform several simulations at the same $\zeta=1.264$ for rings with $U=1,000$.
Figure \ref{fig:D0} displays the MSD as a function of $D_0$ for this state point, clearly showing that the superdiffusive regime is present in all cases. We notice that, while the extension of the time window in which super-diffusion occurs remains fairly constant, it shifts towards larger times upon decreasing $D_0$.

\bibliography{BiblioCircles}

\end{document}